\documentstyle[a4]{article}

\begin{document}

\newcommand{\bib}{\bibitem}
\newcommand{\er}{\end{eqnarray}}
\newcommand{\br}{\begin{eqnarray}}
\newcommand{\be}{\begin{equation}}
\newcommand{\ee}{\end{equation}}
\newcommand{\epe}{\end{equation}}
\newcommand{\bea}{\begin{eqnarray}}
\newcommand{\eea}{\end{eqnarray}}
\newcommand{\ba}{\begin{eqnarray}}
\newcommand{\ea}{\end{eqnarray}}
\newcommand{\epa}{\end{eqnarray}}
\newcommand{\ar}{\rightarrow}
\def\I{{\cal I}}
\def\A{{\cal A}}
\def\F{{\cal F}}
\def\a{\alpha}
\def\b{\beta}
\def\r{\rho}
\def\D{\Delta}
\def\R{I\!\!R}
\def\l{\lambda}
\def\d{\delta}
\def\T{\tilde{T}}
\def\k{\kappa}
\def\t{\tau}
\def\f{\phi}
\def\p{\psi}
\def\z{\zeta}
\def\G{\Gamma}
\def\ep{\epsilon}
\def\hx{\widehat{\xi}}
\def\na{\nabla}
\newcommand{\bslash}{b\!\!\!/}
\newcommand{\vslash}{v\!\!\!/}
\newcommand{\eslash}{e\!\!\!/}
\newcommand{\rslash}{r\!\!\!/}
\begin{center}

{\bf The Kalb-Ramond field as a connection on a flat space time.}

\vspace{1.3cm} M. Botta Cantcheff\footnote{e-mail: botta@cbpf.br}

\vspace{3mm} Centro Brasileiro de Pesquisas Fisicas (CBPF)

Departamento de Teoria de Campos e Particulas (DCP)

Rua Dr. Xavier Sigaud, 150 - Urca

22290-180 - Rio de Janeiro - RJ - Brazil.
\end{center}

\begin{abstract}

We obtain a (Abelian) two form field as a connection on a flat
space-time and its corresponding field strength is canonically
constructed.

\end{abstract}

\section{Introduction.}

The so-called (Abelian) Kalb-Ramond field \cite{kr0,kr},
$B_{\mu\nu}$, is a two-form field which appears in the low-energy
limit of String Theory and in several other frameworks in Particle
Physics \cite{aplic}; for instance, most of the attempts to
incorporate topological
 mass for the field theories in four dimensions take in account this object \cite{tm0,tm}.
  Its dynamics is governed by an action which remains invariant under
transformations whose form are extremely similar to those of a
one-form gauge field \cite{la}. This discussion is essential, for
instance, to understand how the KR-field (which arises naturally
in String Theory) must interact with the matter fields in a gauge
invariant way and if it is associated to new conserved charges.

 The Kalb-Ramond (KR) field transforms according to the following rule: \be B_{\mu\nu} \to
B_{\mu\nu} + \partial_{[\mu}\b_{\nu]}, \label{trKR} \ee where
$\b_\nu$ is a one-form parameter. The question is: may this be
systematically generated from a group transformation? In other
words, how can we associate the parameter $\b_{\mu}$ to the
manifold of some gauge group? \cite{otro,ultimo}.  The main
purpose of this letter is to provide an answer to this question,
by completing the construction started in ref. \cite{gkr}, where a
fundamental step in order to determine the structure associated
with this symmetry was given: The gauge parameter, $\b$, results
to be a tensor product of an algebra element and a one-form, which
 needs to be {\it fixed}.
 This separability condition
for the Lie parameter is somewhat unexpected, since it does not
seem to be a restriction arising from the transformation
properties of a KR-field. Here we shall show that, when one
considers tensorial products of these groups,
 this condition is removed, the total KR-gauge parameter is no longer separable
and the connection is a rank-two tensor field.

This work is organized as follows: in the Section 2, we construct
a Lie group such that a two form field may be recovered from the
connection; finally in Section 3, the covariant derivative and the
Field strength for an Abelian Lie group are defined and the
two-form field appears in the connection.

\section{The group structure for a two-form connection.}

Let us assume a four-dimensional oriented Minkowski space-time
$(M,\eta_{\mu \nu}(\equiv diag[1,1,1,-1])$ and some Lie group
denoted by $G$ whose associated algebra is ${\cal G}$;
${\tau}^{a}$ are the matrices representing the generators of group
with $a= 1,\ldots , \mbox{dim}\:G$; $\tau_{abc}$ are the structure
constants. Our gauge parameter is \be \b_\mu = \b^a_\mu \tau^a \ee
Consider also a Dirac's spinor space ${\cal S}=\{\psi_I\}$ as a
representation for $G$, where $I,J=1..4$ denotes the spinor
components\footnote{Actually, $\psi_I$ is an $N$-component spinor,
where $N$ is the dimension of the group representation.}.

As it has been argued in our previous article \cite{gkr}, this
parameter needs to be separable: $\b^a_\mu =\b^a\,v_\mu$; where
$v_\mu \in \Lambda_1$ is {\it fixed} and $\b^a \in {\cal G}$. So,
 for each fixed direction in $\Lambda_1$, one have a
parametrization of the Lie group. A natural scalar (algebra
valued), $\alpha$, must be introduced and the general form of a
group parameter is $\rho \equiv ( \alpha^a I + \b^a_{\mu}(= \b^a
v_{\mu}) \gamma^\mu )\tau^a$ for a $v_\mu$ fixed. $\{\gamma^\mu
\}_{\mu=1}^4$ are the Dirac's matrices which satisfy the well
known algebra: $\{\gamma^\mu , \gamma^\nu \}_{IJ} = \eta^{\mu \nu}
I_{IJ}$ \footnote{Therefore $\{\gamma^i \}_{i=1}^3$ are hermitian
while $\gamma^4$ is anti-hermitian; thus, if $\tau^a$ are
hermitian and $v$ is time-like, $\rho^a$ must thus be
anti-hermitian and $\a^a$, purely imaginary numbers. The group
manifold one is considering in this case is the one obtained from
$G$ by replacing the Lie parameter $\alpha$ by $i \alpha$. The
unitarity or not of the groups is not determinant for our
discussion; however, in order to avoid this type of considerations
one may take an Euclidean space-time metric. }. The parameter
$\rho$ is an element of the Lie algebra: ${\bar{\cal G}}={\cal G}
\oplus {\cal G}$, the direct sum of two copies of ${\cal G}$
associated with the matrices $I$ and $\vslash$ respectively. The commutators
algebra is
 given by: \be [\r , \r' ]= [\a , \a'] + [\b , \b'] v^2 + ([\a, \b'] + [\b , \a'] ) \vslash\ee

\vspace{0.7cm}

However, we do not need to repeat here the arguments of ref. \cite{gkr}
since, for our present aim, we will only use that \be
\label{resultado} g(\alpha , \beta)= \exp{i\left( \alpha^a + (\b^a
v_{\mu}) \gamma^\mu \right) \tau^a} \ee is a well-defined
(generic) representative of a element of the Lie group ${\bar G}$
which results from the exponentiation of ${\bar{\cal G}}$. In the
Abelian case this simply reduces to  ${\bar G}_{(v)} \sim G_{(I)}
\circ G_{(\vslash)}$, the {\it composition} of two copies of $G$
\footnote{We stand for ${\bar G}_{(v)}$, the group associated with
the space-time direction, $v$.}.

The question now is: how may we define the covariant derivative in
such a way that the KR field be obtained in the connection? This
is the main purpose of this letter.

Notice that the group formed by the collection of the elements
(\ref{resultado}) is not enough to recover the KR field from the
corresponding connection: as usual, one would have a one-form
connection $A_\mu^{(v)}$, associated with the direction $v$ in the
space-time; {\it however}, the transformation law (Abelian for
simplicity) would be $\delta A_\mu^{(v)} = \partial_\mu \b^{(v)}
$. This reveals that $A_\mu^{(v)}$
 {\it is not} a one-form
but a component of a {\it rank-two} tensor field; since the gauge
parameter $\b^{(v)}$ is not a scalar but
 a component of a one form.

In addition, it is clear that a two-form may be thought as a {\it
collection} of four one-forms; each one associated with an element
of an orthonormal basis of the space-time, i.e \be B_{\mu\nu} \sim
\{ B_{\mu}^{\underline{\nu}} \}_{\underline{\nu}}.\ee

The underlined ${\underline{\mu}}$ manifestly refers only to the
component $\mu$ of a tensor (Ex. $\langle dx^{\underline{\mu}} ; V
\rangle = V^{\underline{\mu}} $ for a vector $V$), where $\{
dx^{\underline{\mu}} \}_{{\underline{\mu}}=0}^3 $ is a cartesian
(coordinate) basis for $\Lambda_1$ \footnote{An {\it inertial}
orthonormal frame on Minkowski space-time.}; furthermore,
 repetitions of these indices do not
indicate summation over.

This point of view induces us to propose a group element as being
a {\it tensor product} of (four) elements of the respective
groups, ${\bar G}_{\underline{\mu}}\equiv {\bar G}_{v \equiv
dx^{\underline{\mu}}}$:

  \be\label{tens-prod} g \equiv
\otimes^4_{\mu=1} \left( g_{\underline{\mu}} \in
G_{\underline{\mu}} \right),\ee where $ g_{\underline{\mu}}$ is a
well defined group element associated with the direction
$dx^{\underline{\mu}}\in \Lambda_1$, i.e \be
g_{\underline{\mu}}=\exp{i \left( \alpha_{(\underline{\mu})}^a +
\b^a_{\underline{\mu}} \gamma^{\underline{\mu}} )\right) \tau^a},
\ee where $\alpha_{(\underline{\mu})}$ is an algebra valued scalar
parameter and $\b_{\underline{\mu}}dx^{\underline{\mu}}$ is a one
form pointing in the direction of $dx^{\underline{\mu}}$; if we
define the generic one form: $\b \equiv \sum_{\underline{\mu}=1}^4
\b_{\underline{\mu}}dx^{\underline{\mu}}$, we have
 $\b_{\underline{\mu}}=\langle \b ;
\frac{\partial}{\partial x^{\underline{\mu}}}\rangle$ (where
$\langle dx^{\mu} ; \frac{\partial}{\partial x^{\nu}}\rangle
=\delta^\mu_\nu $). We may denote this group as
 $ G \sim \otimes^4_{\mu=1}
G_{\underline{\mu}}$.

In what follows, we restrict to ourselves to Abelian groups. Since
for general (non-Abelian) Lie groups some technical details must
be particularly discussed, we shall analyze this separately.

Next, in order to re-express $g$ in a more convenient form for
algebraic manipulations, we need to define some extra structure.
 The ordering of the cartesian basis implies that we univocally may
define the extensions \be\label{conv0} \Gamma^{\underline{\mu}}
\equiv \underbrace{I\otimes \dots \otimes I}_{\mu-1}\otimes
\gamma^{\underline{\mu}}\otimes\underbrace{I\otimes \dots \otimes
I}_{4-\mu}\; , \ee where $I$ is the identity on the spinor space
${\cal S}$. Notice that $\Gamma^{\underline{\mu}}$ consists in
inserting $\gamma^{\underline{\mu}}$ in the position
$\underline{\mu}$ between tensor products of the identities.

Thus, let us adopt the notation: \be \label{conv1}
\b_{\underline{\mu}} \Gamma^{\underline{\mu}} \equiv
\underbrace{I\otimes \dots \otimes I}_{\mu-1}\otimes
\b_{\underline{\mu}}\gamma^{\underline{\mu}}\otimes\underbrace{I\otimes
\dots \otimes I}_{4-\mu},\ee then, \be\label{conv2} \b_\mu \G^\mu
= \sum_{\mu=1}^4 \b_{\underline{\mu}} \Gamma^{\underline{\mu}} \ee
for an arbitrary one-form $\b_{\mu}$
Given an {\it ordered} basis of the space time,
$\{\frac{\partial}{\partial x^{\mu}} \}_{\mu=0}^3 $ and a one
form, $\b_{\mu}$, the map: $\b_\mu \longmapsto  \b_\mu \G^\mu$, is
univocally defined.
 Let us define also the extended identity \be {\cal I}
\equiv \underbrace{I \otimes \dots \otimes I}_{4} .\ee
 Then, from (\ref{tens-prod}) the
generic group element may be written: \be g (\alpha , \b) = \exp{i
\left( \alpha \I + \b_\mu \Gamma^\mu \right)},\ee where, all the
parameters $\alpha_{(\underline{\mu})}$ have been absorbed in a
only scalar $\alpha$ \footnote{This is verified by using the
property of tensor products among the identities of the different
spaces: \be \label{propI} a\, \I = aI \otimes  I \otimes I\otimes
I = I \otimes a I \otimes I\otimes I = I\otimes I \otimes a \, I
\otimes I = I\otimes I \otimes I \otimes a I = \I a \ee for all
scalar $a$.}.

A crucial property of these objects is the linear independence:
\be\label{li} \alpha \I + \b_\mu \G^\mu =0 \Longrightarrow
\alpha=\b_\mu =0 .\ee

\section{The Abelian Kalb-Ramond Field as a Connection.}

 Due to (\ref{conv0})
and the properties of the tensorial product, we get the
commutation
 relations \be\label{conm}
[\G^\mu ; \G^{\nu}]=0 , \ee where it is crucial to notice the
position of $\gamma^\mu$ in the tensor product of eq.
(\ref{conv0}) determines the commutation  relation above.

In the Abelian case, a group element may be separated as \be g
(\alpha , \b) = \exp{i \alpha \I} ~ \exp{ i \, \b_\mu \Gamma^\mu
}. \ee The factor $e^{i\alpha}$ is not interesting for us because
this generates the usual one form connection and the Maxwell field
strength. Thus, for more simplicity, we drop it out from the
construction presented below.

We define the covariant derivative as \be \nabla_\mu \equiv
\partial_\mu - i B_{\mu}\ee acting on a spinor $\Psi \in
\otimes_\mu \; {\cal S}_{(\mu)} \equiv {\cal S}_4$. Thus, by
assuming that under a gauge transformation $\Psi \to \Psi' = g(\b)
\Psi = e^{i \b_\mu \G^\mu} \Psi $, its
 covariant derivative transforms in according with
  $\nabla'_\mu \Psi'  = g(\b) \nabla_\mu \Psi $,
and $\nabla'_\mu = \partial_\mu - i B'_{\mu}$,we get that the
connection transforms according to \be \label{transB} B'_{\mu}=
e^{-i \b_\nu \G^\nu }( B_{\mu} +
\partial_\mu \b_\nu \, \G^\nu  )e^{i \b_\nu \G^\nu }.\ee Furthermore, for a coordinate basis in
 a flat space-time, we have the property
$\partial_\mu (dx^\nu)=0$, in virtue of this property and
(\ref{conm}), we obtain
 the transformation law for the connection reduces to
\be B'_{\mu} - B_{\mu} = (\partial_\mu \b_\nu)\, \G^\nu \ee as
expected.

This means that the manifest form of the connection is
$B_{\mu}=B_{\mu\nu} \G^\nu $, since it
 must be an object whose nature is preserved under gauge transformation.
Of course, this connection may be decomposed in its symmetric and
anti-symmetric part, as $b_{\mu\nu}\equiv B_{[\mu\nu]}$ and
$G_{\mu\nu}\equiv B_{(\mu\nu)}$. We identify the Kalb Ramond gauge
(Abelian) field with $b_{\mu\nu}$.

\vspace{0.6cm}

Now, we define the field strength for the $B$-connection as usual
by :\be [\nabla_\mu , \nabla_\nu] \Psi = -i  \F_{ \mu \nu} \Psi \;
; \ee we thus obtain \be \F_{ \mu \nu}= 2\partial_{[\mu }
B_{\nu]}\, .\ee Using that \be\F_{ \mu \nu}= \F_{\mu \nu \r} \G^\r
\, , \ee we get the final result: \be \F_{ \mu \nu \r} =
2\partial_{[\mu } B_{\nu] \r}\,.\ee

It is very useful to define the totally antisymmetric part of this
tensor, $H_{ \mu \nu \r} \equiv \F_{ [\mu \nu \r] }$, because the
commonly studied Abelian gauge actions (for instance, the
so-called Cremer-Sherk-Kalb-Ramond model\cite{kr0,tm0}) are
quadratic in the dual of the field strength, $\ep^{\sigma \mu \nu
\r} \F_{2 \mu \nu \r} = \ep^{\sigma \mu \nu \r} H_{\mu \nu \r}$;
notice also that $H$ only involves derivatives of Kalb-Ramond
field $b$, the antisymmetric component of $B$.

Finally, let us mention that this construction may be expressed in
a more covariant notation: let us define the matrices
\be\label{cov1} \Gamma^{\mu}_{(i)} \equiv \underbrace{I\otimes
\dots \otimes I}_{i-1}\otimes
\gamma^{\mu}\otimes\underbrace{I\otimes \dots \otimes I}_{4-i}\; ,
\ee which consists in inserting $\gamma^\mu$ in the position
$i=1,2,3,4$ (note that here, the vectorial index $\mu$ does not
appear underlined since these matrices are not referred to specific basis
elements). These matrices are manifestly independent of the
reference system; under Lorentz transformations, they transform as:
 $\Gamma'^{\mu}_{(i)}=\Lambda^\mu_\nu \,\sigma(\Lambda) \Gamma^{\nu}_{(i)}
\sigma^{-1}(\Lambda)=\Gamma^{\mu}_{(i)}$ (Pauli theorem), where
$\sigma(\Lambda)\equiv S(\Lambda) \otimes S(\Lambda)\otimes
S(\Lambda) \otimes S(\Lambda)$, and $S$ is the usual unitary
transformation induced by the Lorentz transformation
($\Lambda^\mu_\nu$) on the Dirac spinor space.

 Notice that a group
element (\ref{resultado}), corresponds to the Lorentz transformed,\\
$g^{\Lambda}(\alpha , \beta)= \exp{i\left( \alpha^a + S (\b^a
v_{\mu}) \gamma^\mu ) S^{-1}\right) \tau^a}$, in the
representation ${\cal S}^{\Lambda} = \{ S(\Lambda) \Psi \; / \, \Psi \in {\cal S} \}$.

On the other hand, given a particular orthonormal basis of the
space-time\\
 $\{e^{\mu}_{(i)}\}_{i=1}^4$ (such that $\partial_\a
e^{\mu}_{(i)}=0 $), there exists a unique way of writing an
arbitrary one-form: \be \b_\mu = \sum_i \b_{\mu}^{(i)}
~~~~~(\Lambda^1 = \oplus_{(i)} \Lambda^{1}_{(i)}) . \ee Thus,
for all one form $\b_\mu$, the group parameter, \be {\cal B}
\equiv \sum_i \b_{\mu}^{(i)} \Gamma^{\mu}_{(i)},\ee is uniquely
defined. This is a geometrical object which transforms as: ${\cal
B}\to \sigma {\cal B}\sigma^{-1}$.

 The non-Abelian extension
 of this structure may be built up by following this thought line
where supplementary one-form gauge fields must be introduced in
the connection, a satisfactory fact in the context of duality and
BF-type theories \cite{dob}.
 In a forthcoming paper, we are going to show this and to study diverse gauge theories
including the coupling with fermionic matter that may naturally be
constructed in this new framework \cite{teorkr}.

{\bf Aknowledgements}: The author is specially indebted to Alvaro
L. M. A. Nogueira for many invaluable discussions, relevant
comments and criticisms. Special thanks are due to Prof. A. Lahiri
for pertinent and estimulating comments on the approach developed here.
 Prof. S. Alves D\'{\i}as is
acknowledged for discussions and  Prof.
J. A. Helayel-Neto for clarifyng technical questions on Group
Theory and for the helpful corrections on the original manuscript.

\end{document}